\pdfoutput=1

\documentclass[3p,times,twocolumn]{elsarticle}

\usepackage{ecrc}

\volume{00}

\firstpage{1}

\journalname{Nuclear Physics B Proceedings Supplement}

\runauth{D. Boncioli, A. di Matteo and A.F. Grillo}

\jid{nuphbp}

\jnltitlelogo{Nuclear Physics B Proceedings Supplement}

\usepackage{amssymb}

\usepackage{lineno}

\biboptions{compress}

\usepackage[figuresright]{rotating}

\usepackage{amsmath}
\usepackage{hyperref}
\newcommand{\EeV}{\mathrm{EeV}}
\newcommand{\eV}{\mathrm{eV}}
\newcommand{\gCMB}{\gamma_\text{CMB}}
\newcommand{\p}{\mathrm{p}}
\newcommand{\n}{\mathrm{n}}
\newcommand{\e}{\mathrm{e}}
\newcommand{\arXiv}[1]{\href{http://arxiv.org/abs/#1}{\nolinkurl{arXiv:#1}}}
\newcommand{\doi}[1]{\href{http://dx.doi.org/#1}{\nolinkurl{doi:#1}}}
\newcommand{\mail}[1]{\href{mailto:#1}{\nolinkurl{#1}}}
\newcommand{\Xmax}{X_{\max}}
\newcommand{\SimProp}{\textit{SimProp}}
\newcommand{\Einj}{E_\text{inj}}
\newcommand{\Rcut}{R_\text{cut}}
\begin{document}

\begin{frontmatter}



\dochead{}

\title{Surprises from extragalactic propagation of UHECRs}


\author[lngs]{Denise Boncioli\fnref{DESY}}
\fntext[DESY]{Now at DESY, Zeuthen, Germany.}
\author[univaq]{Armando di Matteo\corref{speaker}}
\cortext[speaker]{Speaker.}
\ead{armando.dimatteo@aquila.infn.it}
\author[lngs]{Aurelio Grillo}
\author[auger]{for the Pierre Auger Collaboration\corref{list}}
\ead{auger\_spokespersons@fnal.gov}
\cortext[list]{\raggedright Full author list: \url{http://www.auger.org/archive/authors_2015_09.html}}
\address[lngs]{INFN, Laboratori Nazionali del Gran Sasso, Assergi (L'Aquila), Italy}
\address[univaq]{INFN and Department of Physical and Chemical Sciences, University of L'Aquila,
L'Aquila, Italy}
\address[auger]{Observatorio Pierre Auger, Av.\ San Mart\'in Norte 304, 5613 Malarg\"ue, Argentina}

\begin{abstract}
Ultra-high energy cosmic ray experimental data are now of very good statistical significance even in the region of the expected GZK feature. The identification of their sources requires sophisticate analysis of their propagation in the extragalactic space. When looking at the details of this propagation some unforeseen features emerge. We will discuss some of these ``surprises''.
\end{abstract}




\end{frontmatter}


\section{Introduction}
\subsection{History of UHECR propagation studies}
The story of cosmic ray research began in 1911, when V.F.~Hess discovered that
the radiation causing air ionization must have an extraterrestial origin~\cite{bib:Hess}. Since the
early 1960s, ultra-high energy cosmic rays (UHECRs) have been detected with energies up to~$100~\EeV$ and beyond.
In 1965, A.A.~Penzias and R.W.~Wilson accidentally discovered the cosmic microwave background (CMB)~\cite{bib:CMB},
confirming the Big Bang model, which had been proposed by G.~Lema\^itre in 1931~\cite{bib:BigBang} but
was at first met with skepticism. The next year, K.~Greisen~\cite{bib:Greisen}, G.T.~Zatsepin and V.A.~Kuz'min~\cite{bib:ZK}
predicted that the CMB set a limit~$E_\text{GZK} \approx 50~\EeV$ to the energy with which protons from distant
sources can reach us, due to photohadronic processes such as
$\p + \gCMB \to \p + \pi^0$~or~$\p + \gCMB \to \n + \pi^+$
in which such protons would lose large fractions of their energy until back below~$E_\text{GZK}$.
In 1998, the AGASA collaboration announced that their detector array had measured the energy spectrum
of cosmic rays to be a power law with no cutoff up to~$2\times10^{20}~\eV$~\cite{bib:AGASA}, spurring speculation
about mechanisms that could allow protons to elude the GZK limit, but all later experiments,
HiRes~\cite{bib:HiRes}, the Pierre Auger Observatory~\cite{bib:Auger}, and the Telescope Array~\cite{bib:TA}, do see a cutoff in the vicinity of~$E_\text{GZK}$.

\subsection{Protons or nuclei?}
\begin{figure}
  \centering
  \includegraphics[width=0.90\columnwidth]{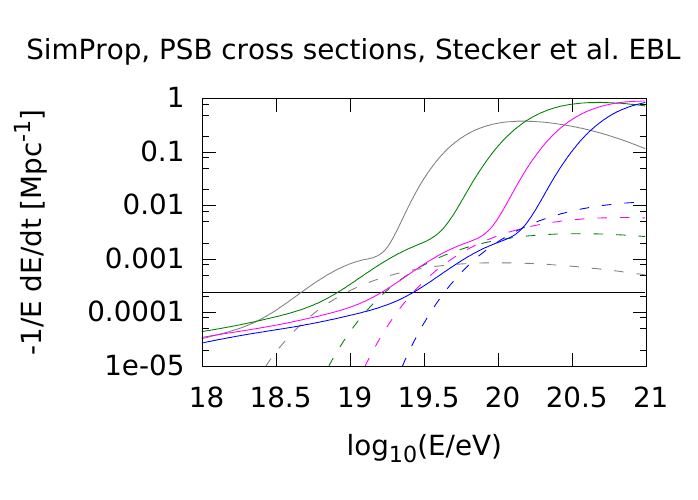}
  \caption{Expected energy loss rate from the expansion of the Universe~(black), photodisintegration~(solid) of helium~(grey), nitrogen~(green), silicon~(magenta) and iron~(blue), and electron--positron pair production~(dashed) for the same elements~(same colours)}
  \label{fig:lambda_nuclei}
\end{figure}

Pierre Auger Observatory data about shower maximum depth ($\Xmax$) distributions,
if interpreted according to models of hadronic interactions in the atmosphere tuned
to LHC data, suggest that most cosmic rays with~$E\gtrsim 10^{19.5}~\eV$ are nuclei
with mass numbers~$A \approx 14$ with few if any protons~\cite{bib:AugerXmax}. If this is the case, interactions with
background radiation become even more important, because, while the pion production
threshold is shifted to $A$~times that for protons, photodisintegration interactions
(e.g.~${^A Z} + \gamma \to {^{A-1} Z} + \n$) are possible with both CMB 
and extragalactic background light (EBL) photons, with very short interaction lengths even at moderate energies (see Fig.~\ref{fig:lambda_nuclei}).

(The Telescope Array collaboration continues to interpret their $\Xmax$~distributions as being compatible with a largely protonic composition at all energies~\cite{bib:TAXmax}, but in the following we only discuss Auger data because proton-only scenarios have already been extensively studied and are outside the scope of this work. A direct comparison of $\Xmax$~measurements by the two observatories taking into account the different fiducial cuts applied found no incompatibility between them~\cite{bib:Xmaxcomp}.)

\section{Modelling issues}
\subsection{Monte Carlo simulation codes}
Several Monte Carlo codes have been developed to simulate the propagation of
UHECRs. The phenomena
they take into account include the adiabatic energy loss due to the expansion of
the Universe, the electron--positron pair production~$N + \gCMB \to N + \e^+ + \e^-$,
the photodisintegration of nuclei, and pion production.
Two publicly available such codes are \SimProp\footnote{Available upon request to~\mail{SimProp-dev@aquila.infn.it}.}~\cite{bib:tesiDenise,bib:SimProp20,bib:SimProp21,bib:SimProp22,bib:nupaper,bib:SimProp23} and CRPropa\footnote{Available for download from~\url{http://crpropa.desy.de/}.}~\cite{bib:CRPropa1,bib:CRPropa2,bib:CRPropa3}, which are compared in Ref.~\cite{bib:SALpropa}

\subsection{Poorly known quantities (photodisintegration and EBL) and their effects}
Some of the quantities relevant to the propagation of UHECRs
are known with good accuracy; these include the expansion of the Universe (described by the FLRW metric),
the spectrum and evolution of the CMB (described by Planck's law for a black body with temperature~$T\propto1+z$),
cross sections for pair production (described by the Bethe--Heitler formula) and pion production
(measured in collider experiments over a wide range of energies), and total photodisintegration cross sections for certain nuclides.

\begin{figure}
  \centering
  \includegraphics[width=0.90\columnwidth]{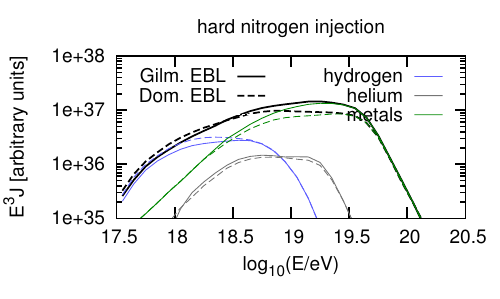}
  \caption{Effect of different EBL models on propagated UHECR fluxes (adapted from Ref.~\cite{bib:SALpropa})}
  \label{fig:EBLeffect}
\end{figure}

\begin{figure}
  \centering
  \includegraphics[width=0.90\columnwidth]{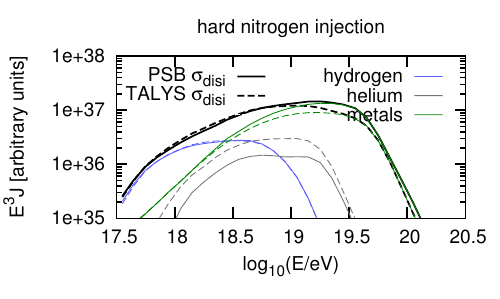}
  \caption{Effect of different photodisintegration models on propagated UHECR fluxes (adapted from Ref.~\cite{bib:SALpropa})}
  \label{fig:sigmaeffect}
\end{figure}

Other quantities are poorly known and need to be approximated via phenomenological models, which do not always agree with the available data.
They include the spectrum and evolution of the EBL and the branching ratios for the various photodisintegration channels.
The effects of the uncertainty in these quantities on results of UHECR propagation simulations are discussed extensively in Ref.~\cite{bib:SALpropa}. Two examples are shown in Fig.~\ref{fig:EBLeffect} and Fig.~\ref{fig:sigmaeffect}, which show that all other things being equal a stronger far-IR peak in the EBL spectrum or larger cross sections for photodisintegration channels in which alpha particles are ejected would result in a softer spectrum and a lighter composition at Earth.

\section{Fitting source models to the data}
The Pierre Auger Collaboration has presented~\cite{bib:fitICRC15} a combined fit to their spectrum~\cite{bib:Auger} and $\Xmax$~\cite{bib:Xmaxscale} data above~$10^{18.7}~\eV$ in a simple astrophysical scenario:
  uniform distribution of identical sources;
  injection consisting of hydrogen-1, helium-4, nitrogen-14 and iron-56 only;
  power-law injection spectrum with broken exponential rigidity cutoff,
  \begin{equation} \frac{\mathrm{d}N_i}{\mathrm{d}\Einj} =
    \begin{cases}
     J_0 p_i \left(\frac{\Einj}{\EeV}\right)^{-\gamma}, & \frac{\Einj}{Z_i} \le \Rcut; \\
     J_0 p_i \left(\frac{\Einj}{\EeV}\right)^{-\gamma}\exp\left(1-\frac{\Einj}{Z_i\Rcut}\right), & \frac{\Einj}{Z_i} > \Rcut. \\
    \end{cases}
  \end{equation}
No hypothesis is made about the origin of the ankle and the UHECR flux below it.

\begin{figure}
  \centering
  \includegraphics[width=0.90\columnwidth]{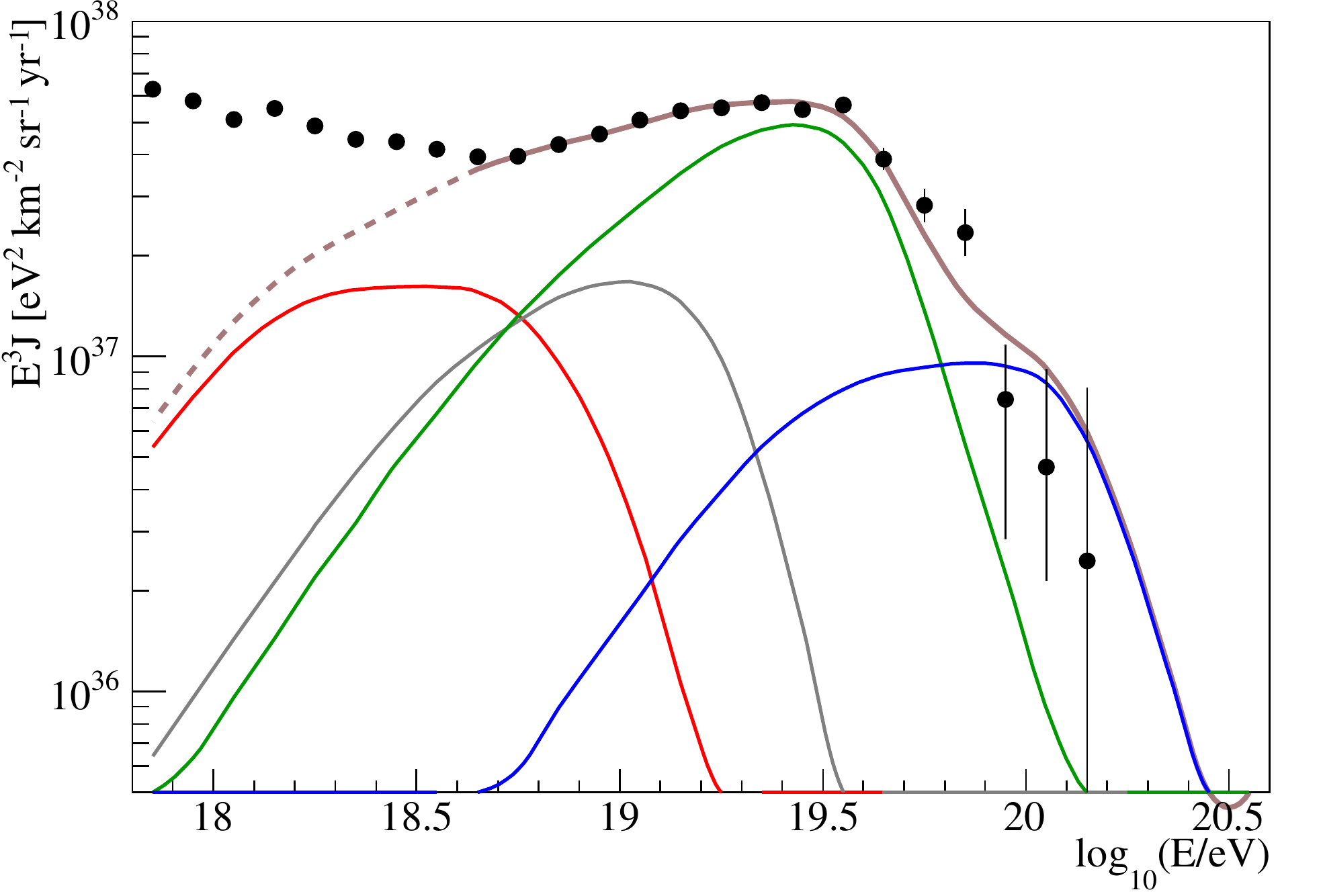}\\
  \includegraphics[width=0.45\columnwidth]{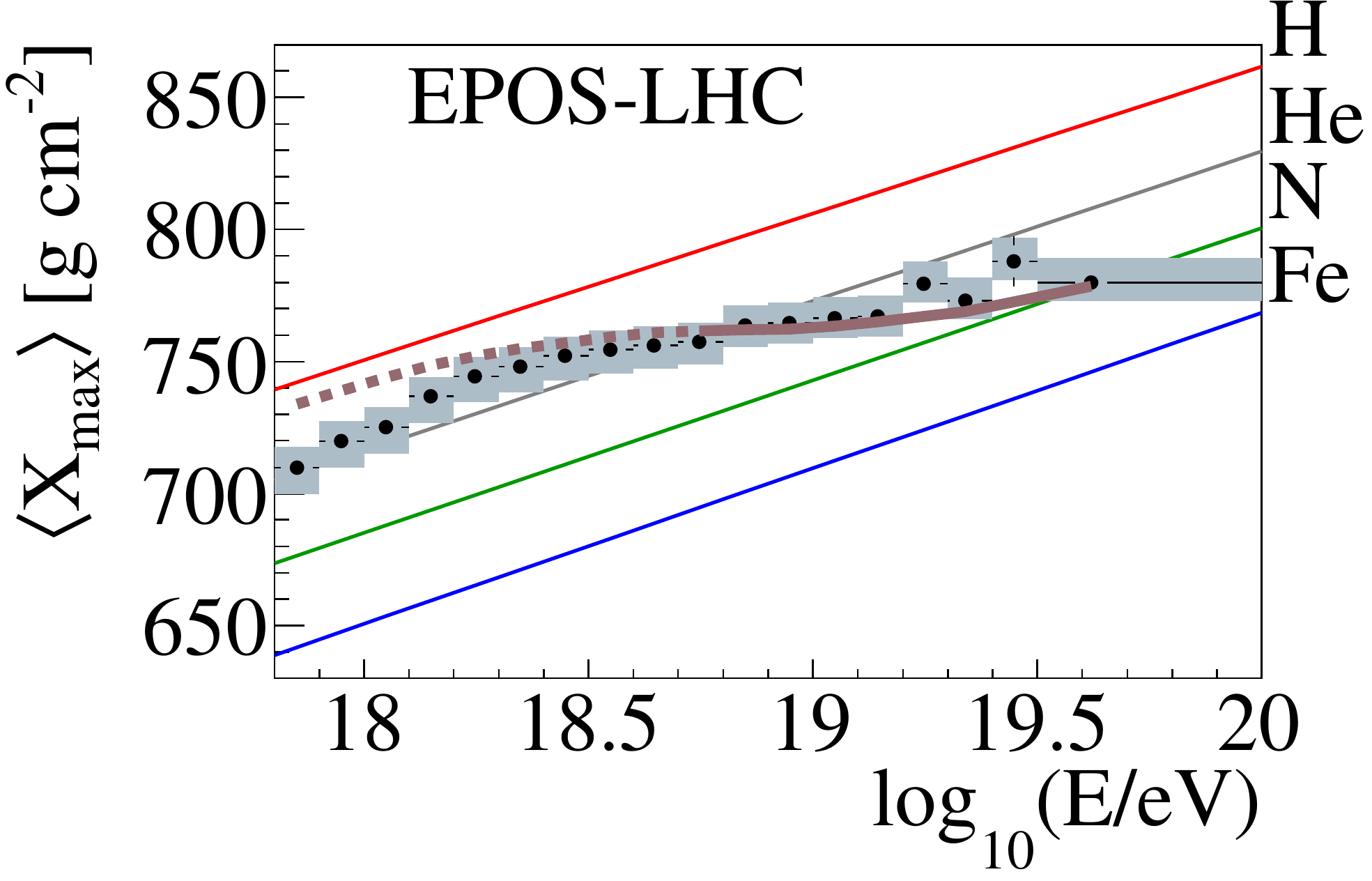}
  \includegraphics[width=0.45\columnwidth]{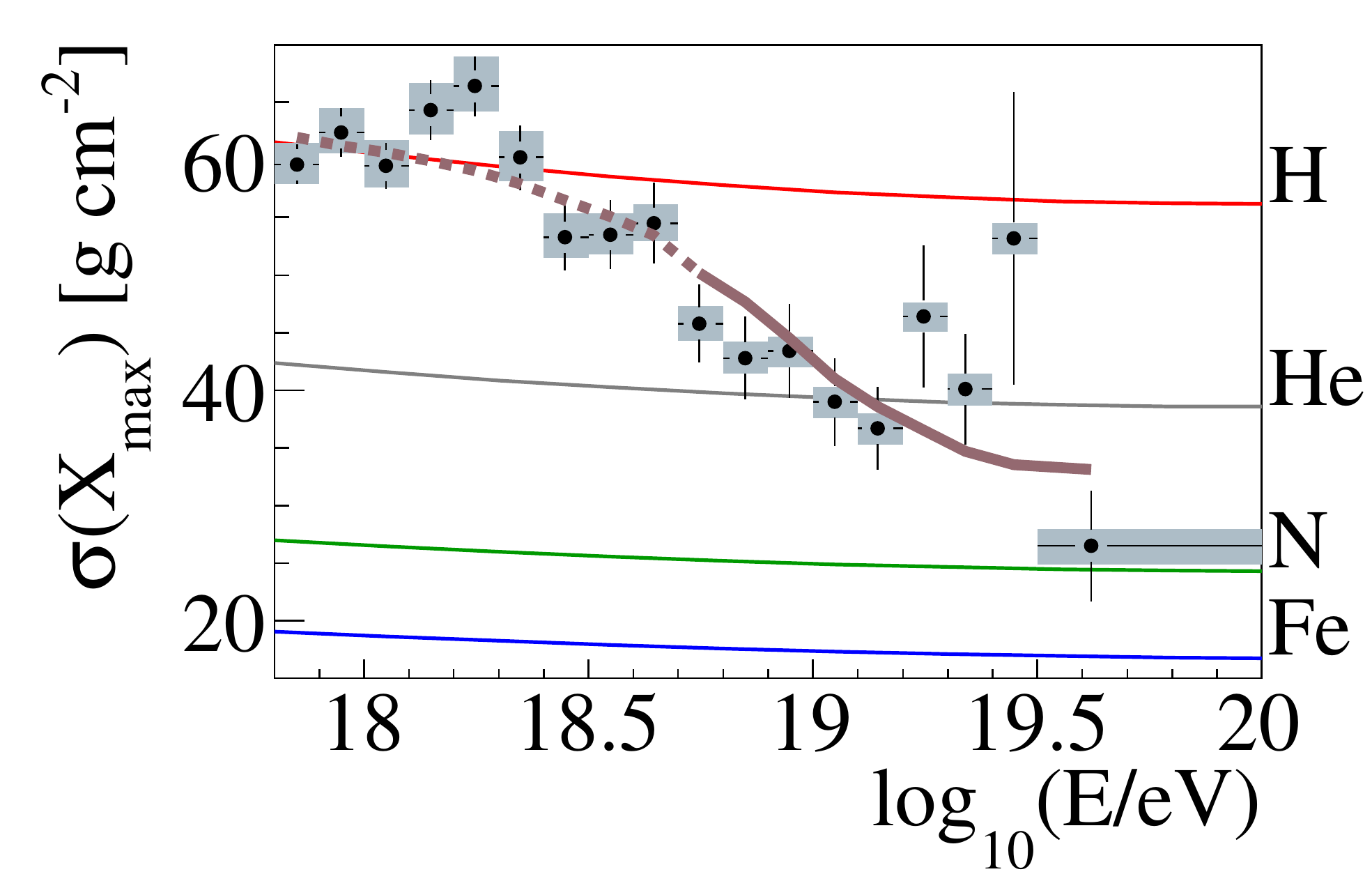}
  \caption{Simulated UHECR fluxes, $\langle\Xmax\rangle$, and $\sigma(\Xmax)$ in the best-fit scenario~\cite{bib:fitICRC15} (thick brown:~total; red:~$A=1$; grey:~$2\le A \le 4$; green:~$5\le A \le 26$; blue:~$A \ge 27$) compared to Auger data~\cite{bib:Auger,bib:AugerXmax}}
  \label{fig:bestfit}
\end{figure}

If the propagation of UHECRs in intergalactic space is simulated with \SimProp~v2r3~\cite{bib:SimProp23} using the Gilmore et~al.~\cite{bib:Gilmore} EBL model and the PSB~\cite{bib:PSB,bib:SS} photodisintegration model and air showers are simulated with CONEX~\cite{bib:CONEX} using the EPOS-LHC~\cite{bib:EPOS} hadronic interaction model, the best fit is found with a relatively hard injection spectral index~$\gamma=0.94^{+0.09}_{-0.10}$, a relatively low cutoff rigidity~$\Rcut = 10^{18.67\pm0.03}~\mathrm{V}$, and a very metal-rich composition $p_\mathrm{H} = 0.0^{+29.9}\%$, $p_\mathrm{He} = 62.0^{+3.5}_{-22.2}\%$, $p_\mathrm{N} = 37.2^{+4.2}_{-12.6}\%$ and $p_\mathrm{Fe} = 0.8^{+0.2}_{-0.3}\%$.
The corresponding simulated spectrum and average and standard deviation of $\Xmax$~distributions are shown and compared to Auger data in Fig.~\ref{fig:bestfit}.
The deviance (generalized~$\chi^2$) of this fit per degree of freedom at the best fit is $D_{\min}/n=178.5/119$, corresponding to a $p$-value of~$2.6\%$.

\begin{figure}
  \centering
  \includegraphics[width=0.90\columnwidth]{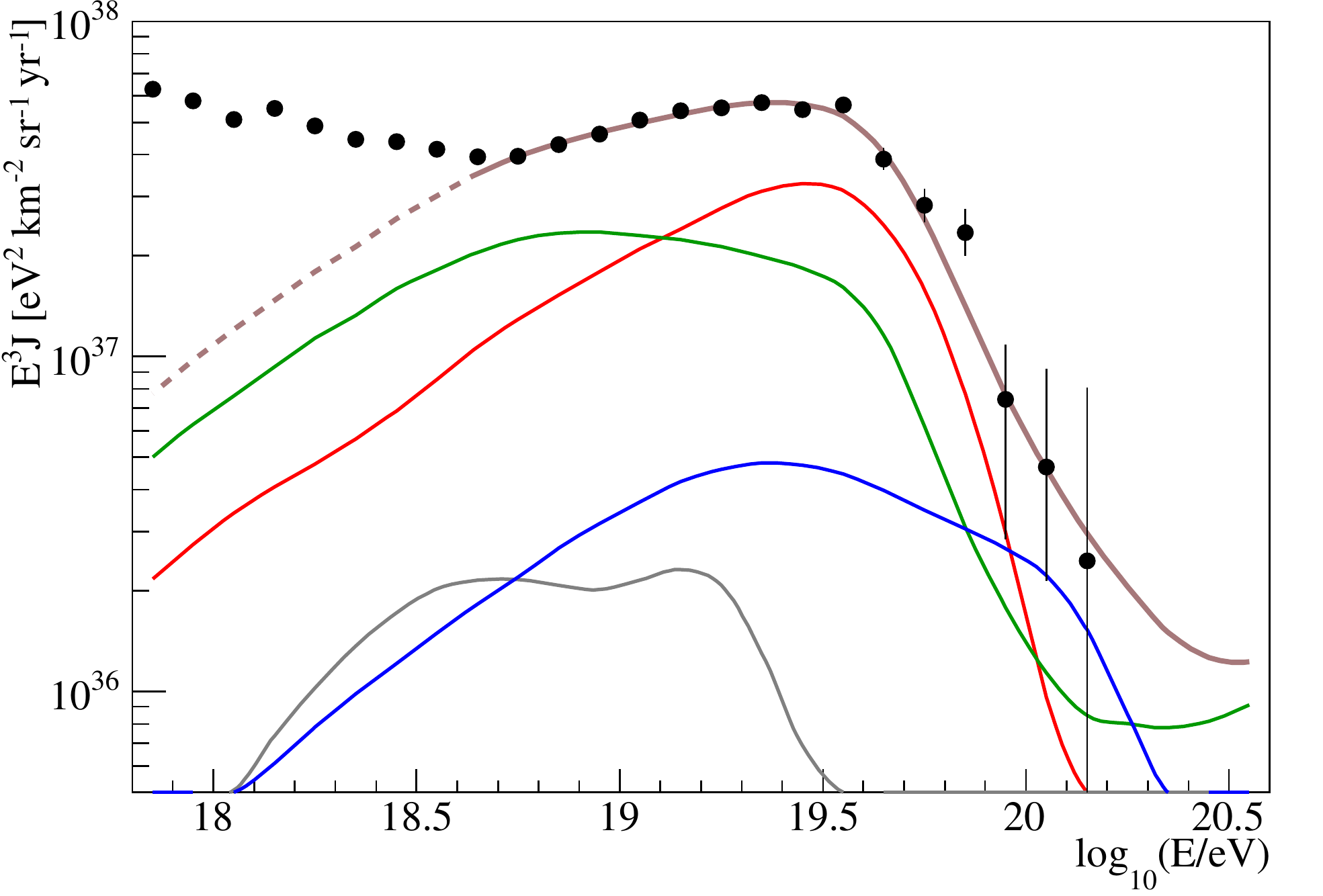}\\
  \includegraphics[width=0.45\columnwidth]{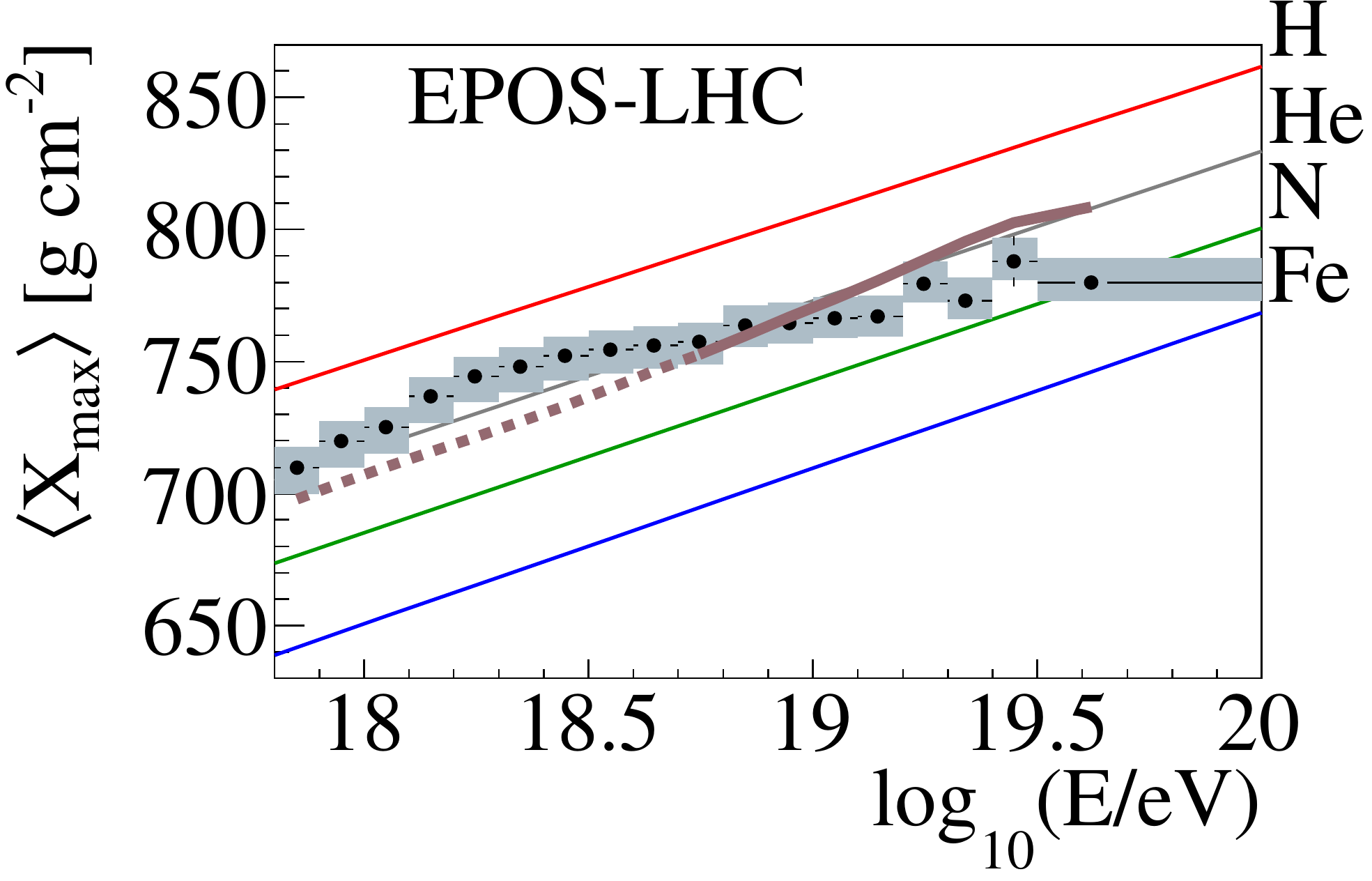}
  \includegraphics[width=0.45\columnwidth]{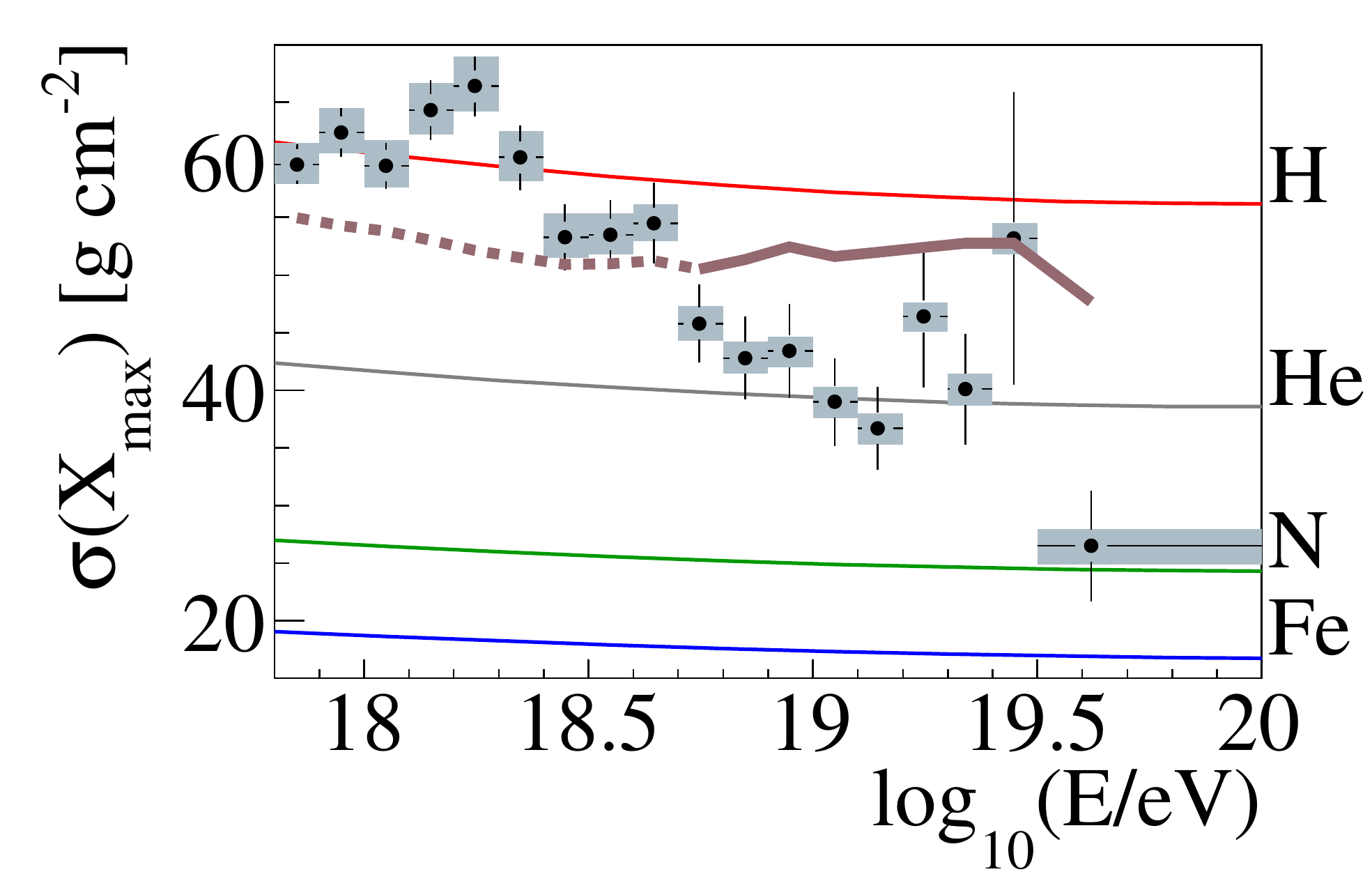}
  \caption{Same as in Fig.~\ref{fig:bestfit} for the second local minimum at~$\gamma\approx2$}
  \label{fig:2ndmin}
\end{figure}

The deviance of the fit as a function of~$\gamma$ and~$\Rcut$ also has a second local minimum at~$\gamma=2.03$, $\Rcut = 10^{19.84}~\mathrm{V}$. The corresponding spectrum and~$\Xmax$ are shown in Fig.~\ref{fig:2ndmin}.
This minimum is disfavoured at the $7.5\sigma$~level compared to that at~$\gamma\approx1$ ($D_2 = 235.0=D_{\min}+56.5$; $p=5\times10^{-4}$), mostly by the observed $\Xmax$~distributions narrower than the predictions at the second local minimum.

In order to assess the dependence of these results on poorly known quantities, the same fit was repeated with different Monte Carlo codes, EBL models, photodisintegration models, air interaction models, and with data shifted by their systematical uncertainties. As shown in Table~\ref{tab:params}, the best-fit parameter values are strongly dependent on the model used, whereas the position of the local minimum at~$\gamma\approx 2$ is much more stable. Also, in most cases models of propagation with lower interaction rates tend to result in better fits to Auger data, as does shifting the energy and~$\Xmax$ scales downwards, and models of hadronic interactions other than EPOS-LHC are strongly disfavoured.
\begin{table*}
  \centering
  \begin{tabular}{@{}cccccc|ccc|ccc@{}}
  \multicolumn{6}{c}{models} & \multicolumn{3}{c}{best fit} & \multicolumn{3}{c}{2nd min} \\
  MC code & EBL & $\sigma_\text{disint.}$ & air int. & $\Delta E/E$ & $\Delta X_{\max}$ & $\gamma$ & cut & $D_{\min}$ & $\gamma$ & cut & $D_2$ \\ 
  \hline
  \textit{SimProp} & Gilmore & PSB & EPOS & 0 & 0 & $+0.94$ & 18.67 & 178.5 & 2.03 & 19.84 & 235.0 \\
  \textit{SimProp} & Dom\'inguez & PSB & EPOS & 0 & 0 & $-0.45$ & 18.27 & 193.4 & 1.98 & 19.77 & 278.7 \\
  \textit{SimProp} & Gilmore & TALYS & EPOS & 0 & 0 & $+0.69$ & 18.60 & 176.9 & 2.01 & 19.83 & 255.7 \\
  CRPropa & Gilmore & TALYS & EPOS & 0 & 0 & $+0.73$ & 18.58 & 195.3 & 2.04 & 19.86 & 253.2 \\
  CRPropa & Dom\'inguez & TALYS & EPOS & 0 & 0 & $-1.06$ & 18.19 & 192.3 & 1.99 & 19.81 & 304.9 \\
  CRPropa & Dom\'inguez & Geant4 & EPOS & 0 & 0 & $-1.29$ & 18.18 & 192.5 & 1.97 & 19.80 & 306.0 \\
  \textit{SimProp} & Gilmore & PSB & EPOS & $-14\%$ & 0 & $+0.90$ & 18.64 & 165.5 & 2.00 & 19.72 & 218.2 \\
  \textit{SimProp} & Gilmore & PSB & EPOS & $+14\%$ & 0 & $+0.98$ & 18.70 & 214.9 & 2.05 & 19.95 & 261.9 \\
  \textit{SimProp} & Gilmore & PSB & EPOS & 0 & $-1\sigma$ & $+1.35$ & 18.73 & 172.1 & 2.08 & 19.86 & 219.7 \\
  \textit{SimProp} & Gilmore & PSB & EPOS & 0 & $+1\sigma$ & $-1.50$ & 18.24 & 217.2 & 2.01 & 19.84 & 241.2 \\
  \textit{SimProp} & Gilmore & PSB & EPOS & $-14\%$ & $-1\sigma$ & $+1.32$ & 18.68 & 157.4 & 2.06 & 19.73 & 204.8 \\
  \textit{SimProp} & Gilmore & PSB & EPOS & $-14\%$ & $+1\sigma$ & $-1.50$ & 18.22 & 207.0 & 2.05 & 19.78 & 234.3 \\
  \textit{SimProp} & Gilmore & PSB & EPOS & $+14\%$ & $-1\sigma$ & $+1.39$ & 18.78 & 203.5 & 2.10 & 19.99 & 235.4 \\
  \textit{SimProp} & Gilmore & PSB & EPOS & $+14\%$ & $+1\sigma$ & $-1.34$ & 18.28 & 256.0 & 2.02 & 19.93 & 274.1 \\
  \textit{SimProp} & Gilmore & PSB & Sibyll & 0 & 0 & $-1.50$ & 18.27 & 256.8 & 2.04 & 19.85 & 293.3 \\
  \textit{SimProp} & Gilmore & PSB & QGSJet & 0 & 0 & $-1.50$ & 18.28 & 344.3 & 2.09 & 19.88 & 317.2 \\
  \end{tabular}
  \caption{Best fit and second local minimum for various combinations of models. The Monte Carlo codes are \textit{SimProp}~v2r3 \cite{bib:SimProp23} and CRPropa~3 \cite{bib:CRPropa3}, the EBL models are Gilmore et~al.~2012 \cite{bib:Gilmore} and Do\-m\'in\-guez et~al.~2011 \cite{bib:DominguezEBL}, the photodisintegration models are PSB~\cite{bib:PSB,bib:SS}, TALYS~\cite{bib:TALYS} and Geant4~\cite{bib:Geant4} (but see Ref.~\cite{bib:SALpropa} for details about the way they are treated in MC codes), the air interaction models are EPOS-LHC \cite{bib:EPOS}, Sibyll~2.1 \cite{bib:Sibyll} and QGSJet~II-04 \cite{bib:QGSJet}, and the shifts on the energy~\cite{bib:Escale} and~$\Xmax$ scales~\cite{bib:Xmaxscale} correspond to $\pm1$~standard deviation. The lowest value considered for~$\gamma$ was~$-1.50$, so models where the best fit is found at~$\gamma=-1.50$ might be improved by lowering~$\gamma$ even further.}
  \label{tab:params}
\end{table*}

\section{Conclusions and future directions}
If interpreted according to a simple astrophysical model of UHECR sources, Pierre Auger Observatory data seem to point to very hard, metal-rich injection spectra. This result is qualitatively similar to recent phenomenological models by other authors \cite{bib:Allard,bib:FKO,bib:ABB,bib:GAP,bib:TAH,bib:UFA}, but is at odds with traditional theoretical models of UHECR acceleration, which predict~$\gamma\gtrsim 2$.

Another important result is that when injection spectra are hard, propagated spectra and compositions are strongly sensitive on poorly known details of interactions with background photons taking place during the propagation. Most studies of UHECR propagation until recently disregarded our ignorance of these quantities. This was acceptable when such studies only considered soft injection spectra ($\gamma \gtrsim 2$) where the differences between various models have no major impact, but now that hard injection spectra ($\gamma \lesssim 1$) are also commonly considered, this is no longer legitimate.

Possible refinements of this astrophysical model that could be made in the near future include considering more than four possible elements at injection, extending the fit to energies below the ankle, and considering non-uniform distributions of sources.







\end{document}